
\input phyzzx

\normalparskip=0pt
\parindent 36pt
\sectionminspace=6pc
\chapterminspace=6pc
\referenceminspace=6pc

\def\unlock{\catcode`@=11} 
\def\lock{\catcode`@=12} 
\unlock
\def\refitem#1{\r@fitem{#1.}}
\def\chapter#1{\par \penalty-300 \vskip\chapterskip
   \spacecheck\chapterminspace
   \chapterreset \leftline{\bf \chapterlabel.~#1}
   \nobreak\vskip\headskip \penalty 30000
   {\pr@tect\wlog{\string\chapter\space \chapterlabel}} }
\def\section#1{\par \ifnum\lastpenalty=30000\else
   \penalty-200\vskip\sectionskip \spacecheck\sectionminspace\fi
   \gl@bal\advance\sectionnumber by 1
   {\pr@tect
   \xdef\sectionlabel{\chapterlabel.
       \the\sectionstyle{\the\sectionnumber}}%
   \wlog{\string\section\space \sectionlabel}}%
   \noindent {\it\sectionlabel.~~#1}\par
   \nobreak\vskip\headskip \penalty 30000 }
\lock

\def\SCIPP{\centerline {\it Santa Cruz Institute for Particle Physics}
\centerline {\it University of California, Santa Cruz, CA 95064}}
\def\quar{{\textstyle{1\over4}}}
\def\shaf{{\textstyle{1\over2}}}

\def\pa{\partial}
\def\Asl{\relax{\rm /\kern-.58em A}}
\def\Bsl{\relax{\rm /\kern-.56em B}}
\def\Dsl{\relax{\rm /\kern-.48em D}}
\def\pasl{\relax /\kern-.56em \pa}
\def\intk{\int {d^3k\over (2\pi)^3}}
\def\dph#1{{\delta {#1}\over\delta\phi}}
\def\mmag{m_{\rm mag}}

\REF\ourpaper{M. Dine, P. Huet, R.G. Leigh, A. Linde and D. Linde,
to appear in {\sl Phys.~Lett.} {\bf B}.}
\REF\ourpapera{M. Dine, P. Huet, R.G. Leigh, A. Linde and D. Linde,
 to appear in {\sl Phys.~Rev.} {\bf D}.}
\REF\kirch{D.A. Kirzhnits, {\sl JETP Lett.} {\bf 15}, 529 (1972);
 D.A. Kirzhnits and A. Linde, {\sl Phys.~Lett.} {\bf 42B}, 471 (1972).}
\REF\Wein{S. Weinberg, {\sl Phys.~Rev.} {\bf D9}, 3357 (1974).}
\REF\furth{ L. Dolan and R. Jackiw, {\sl Phys.~Rev.} {\bf D9},
3320 (1974).}
\REF\kirli{D.A. Kirzhnits and A.D. Linde, JETP {\bf 40}, 628 (1974).}
\REF\first{D.A. Kirzhnits and A. Linde, {\sl Ann.~Phys.} {\bf 101},
195 (1976).}
\REF\kuzm{V.A. Kuzmin, V.A. Rubakov and M.E. Shaposhnikov,
{\sl Phys.~Lett.} {\bf B155}, 36 (1985).}
\REF\barI{L. McLerran, {\sl Phys.~Rev.~Lett.} {\bf 62}, 1075 (1989).}
\REF\barII{L. McLerran, M. Shaposhnikov, N. Turok and
 M. Voloshin, {\sl Phys.~Lett.} {\bf 256B}, 451 (1991).}
\REF\barIII{N. Turok and P. Zadrozny, {\sl Phys.~Rev.~Lett.}
 {\bf 65},  2331 (1990); {\sl Nucl.~Phys.} {\bf B358}, 471 (1991).}
\REF\barIV{M. Dine, P. Huet, R. Singleton and L. Susskind,
 {\sl Phys.~Lett.} {\bf 257B},  351 (1991).}
\REF\barV{A. Cohen, D.B. Kaplan and A.E. Nelson, {\sl Nucl.~Phys.}
 {\bf B349}, 727 (1991).}
\REF\dine{M. Dine, P. Huet and R. Singleton, SCIPP-91/08 (1991).}
\REF\ckn{A. Cohen, D. Kaplan and A. Nelson, {\sl Phys.~Lett.} {\bf 263B},
 86 (1991).}
\REF\meandcollab{P. Huet, K. Kajantie, R.G. Leigh and L. McLerran,
 work in progress.}
\REF\hsu{D. Brahm and S. Hsu, Caltech preprints CALT-68-1705 and
 CALT-68-1762 (1991).}
\REF\shap{M.E. Shaposhnikov, CERN preprint TH.6319/91 (1991) and errata.}
\REF\carr{M. Carrington, Minnesota preprint TPI-MINN-91/48-T.}
\REF\dave{D. Brahm, talk at ITP Santa Barbara, April 1992;
 G. Boyd, D. Brahm, S. Hsu, work in progress.}
\REF\kap{J.I. Kapusta, {\em Finite Temperature Field
 Theory}, (Cambridge University Press, 1989).}
\REF\higher{A.D. Linde, {\sl Rep. Prog. Phys.} {\bf 42}, 389 (1979);
 A.D. Linde, {\sl Phys. Lett.} {\bf 93B}, 327 (1980).}
\REF\washout{M.E. Shaposhnikov, {\sl JETP Lett.} {\bf 44},  465 (1986);
 {\sl Nucl.~Phys.} {\bf B287}, 757 (1987);  {\sl Nucl.~Phys.} {\bf B299},
 797 (1988);  A.I. Bochkarev, S.Yu. Khlebnikov and M.E. Shaposhnikov,
 {\sl Nucl.~Phys.} {\bf B329}, 490 (1990).}
\REF\lep{ALEPH, DELPHI, L3 and OPAL collaborations, as presented by M.
 Davier, Proceedings of the International Lepton-Photon Symposium and
 Europhysics Conference on High Energy Physics, eds.~S. Hegarty, K.
 Potter and E. Quercigh (World Scientific, Singapore, 1991)
 vol.~2, p.~153.}
\REF\bks{A.I. Bockharev, S.V. Kuzmin, M.E. Shaposhnikov,
Phys.~Lett. {\bf 244B}, 275 (1990).}
\REF\tztwo{N. Turok and J. Zadrozny, Nucl.~Phys. {\bf B369}, 729 (1992).}
\REF\andhall{G. Anderson and L.J. Hall, LBL preprint LBL-31169 (1991).}
\REF\ginsparg{P. Ginsparg, {\sl Nucl.~Phys.} {\bf B170}, 388 (1980).}
\REF\LGPY{A.D. Linde, Phys.~Lett. {\bf 93B}, 327 (1980);
 D.J. Gross, R.D. Pisarski and L.G. Yaffe, Rev.~Mod.~Phys. {\bf 53},
 1 (1981).}
\REF\ann{A. Nelson, private communication.}
\REF\erice{S. Coleman, {\em Aspects of Symmetry: Selected Erice lectures}
(Cambridge Univ.~Press, 1985).}
\REF\stein{P.J. Steinhardt, {\sl Phys.~Rev.} {\bf D25}, 2074 (1982).}
\REF\kaj{K. Enquist, J. Ignatius, K. Kajantie and K.  Rummukainen,
 preprint HU-TFT-91-35 (1991).}
\REF\linde{A.D. Linde, {\sl Phys.Lett.} {\bf 70B}, 306 (1977); {\bf
 100B}, 37 (1981); {\sl Nucl.~Phys.} {\bf B216},  421 (1983).}
\REF\Turok{N. Turok, Princeton preprint PUPT-91-1273.}
\REF\lmt{B.-H. Liu, L. McLerran, N. Turok, Minnesota preprint
 TPI-MINN-92-18-T.}

\Pubnum{SCIPP 92/23}
\pubtype{T }     
\date{\today}
\titlepage
\vskip4cm
\singlespace
\title{{\bf INFRARED EFFECTS AND BUBBLE PROPAGATION \break
 AT THE ELECTROWEAK PHASE TRANSITION}
\foot{Work supported in part by the U.S.~Department of Energy.}}
\vskip12pt
\centerline{\caps R.G. Leigh}
\address{\SCIPP}
\vfill
\vbox{ \narrower   
\centerline{ABSTRACT}
We discuss aspects of poor infrared behaviour of the perturbation
expansion for the effective potential for the Higgs mode near
the electroweak phase transition, and enlarge on the discovery that
higher order effects weaken the transition. In addition, we outline our
recent attempts at understanding the dynamics involved in the
propagation of bubbles formed in the first order transition.
}
\vfill
\centerline{Submitted to \it Proceedings of the TEXAS Symposium}
\centerline{\it on Electroweak Baryon Number Violation, Yale University,
March 1992.}
\vfill
\endpage

In this talk, I will discuss several issues related to the theory
of the electroweak phase transition. A discussion of other aspects
which we have recently addressed\refmark{\ourpaper,\ourpapera}\
can be found in the article by A. Linde in these proceedings.

The theory of phase transitions in the early Universe was
pioneered in the seminal work of Kirzhnits and Linde in 1972.%
\refmark\kirch\  Early works claimed that the phase transition
corresponding to the breakdown of SU(2) $\times$ U(1) was of
the second order.\refmark{\Wein,\furth,\kirli}  Finally in 1976, it was
shown\refmark\first\ to be of the first order. This fact
essentially lay dormant for many years
until it was realized\refmark\kuzm\ that this character of the phase
transition can provide the non-equilibrium environment essential
for producing a baryon asymmetry at these scales; the baryon violating
process is typified (at temperatures below the critical temperature)
by the sphaleron. We now believe that the phase transition proceeds
by the nucleation of critical true vacuum bubbles from a rather
weakly supercooled state. These bubbles expand under pressure
forces and quickly fill space. Baryogenesis typically proceeds in
the bubble walls where $\phi$ is changing,\refmark{\barI - \dine}\
or in a thin layer in front of the wall.\refmark\ckn\

It is potentially important for a detailed understanding of the
baryon asymmetry, to understand the propagation of the bubble
wall, namely its size and shape, and its terminal velocity (if any).
In the last half of this presentation, I will discuss our
estimates of these quantities, and compare to other recent analyses.
Work is continuing on refining the analysis and considering
new phenomena that may be important.\refmark\meandcollab\

In the first section of this presentation, I will review the standard
analysis leading to the derivation of the effective potential for
the Higgs mode. It has recently been
emphasized\refmark{\hsu - \dave}\ that higher order
corrections are potentially important because of infrared divergences
and indeed can lead to
the breakdown of perturbation theory in relevant regions of
parameter space. We discuss several approaches to resuming
perturbation theory that soften the poor infrared behaviour.
Finally, we outline by powercounting the
region of parameter space for which the resummed perturbation
theory is good, and comment on what effects the analysis has
on various models of baryogenesis at the electroweak scale.

\chapter{The Electroweak Potential}

\section{One-Loop Analysis}

The standard approach to equilibrium properties of finite temperature
field theories is the imaginary time formalism. In this technique, one
continues to Euclidean spacetime, and compactifies the time coordinate
on a circle of radius $\hbar\beta = T^{-1}$. The resulting path
integral is then identical to a Boltzmann sum over states. The fields
in the theory may be expanded in Fourier series over a discrete
set of Matsubara frequencies
$$\psi(\vec{x},\tau) = \sum_n e^{i\omega_n \tau} \psi_n (\vec{x}).
\eqn\mode$$
Requiring the fields to satisfy (anti-)periodic boundary conditions
on the circle leads to $\omega_n = (2n+1)\pi T$ (fermions) or
$\omega_n = 2\pi nT$ (bosons). In this way, the theory can be
thought of as an infinite set of coupled three-dimensional fields with a
hierarchy of masses $m^2 = \omega_n^2 + m_o^2$ where $m_o$ is the
usual zero temperature mass. As we will see later, this viewpoint will be
of particular value to us. The quadratic part of the action (for
bosons) may be written as
$$S_\beta = \shaf T \sum_n \int d^3 x \;\phi_{-n} \left( -\vec{\nabla}^2
+ \omega_n^2 + m_o^2 \right) \phi_n .\eqn\bosact$$

The effective potential for the Higgs field $\phi$ can be computed
by putting in a source and studying the response of the theory to
this source. At the one loop level, this is equivalent to shifting
$\phi$ by its saddle point value; the familiar result is just the
determinant of quadratic fluctuations. That is, at lowest order,
we can neglect interactions and the effective potential is that
of a collection of free particles interacting with a heat bath at
temperature $T$ with masses given by $\phi$. We have
$$
\eqalign{V_{T,1-loop} (\phi) &= V_0 (\phi) - \shaf T \sum_i
\ln \det \left[ -\vec{\nabla}^2 + \omega_n^2 + m_o^2 (\phi) \right] \crr
 &= V_{0,1-loop} + \left( {T^4\over 2\pi^2} \right)
 \sum_i I_\pm \left( {m_i (\phi)\over T} \right) \cr}
 \eqn\oneloop$$
where
$$I_\mp (y) = \pm \int_0^\infty dx\; x^2 \ln \left( 1 \mp e^{-\sqrt{x^2
+ y^2}} \right) .\eqn\ipm$$
Expanding for small $y$, \ie, $m(\phi) \ll T$, we find the following
potential:
$$V_T (\phi) = D (T^2-T_o^2) \phi^2 - ET\phi^3 + \quar \lambda_T
\phi^4\eqn\vexp$$
where $D, E$ and $\lambda_T$ are determined in terms of the gauge and
Yukawa couplings of the theory and the Higgs self-coupling. We will
assume throughout that the Higgs boson makes only a very small
contribution to the potential, \ie, that it is sufficiently light.
The parameters in the effective potential, Eq.~\vexp , are
given in many previous works, including Refs.~\ourpaper,\ourpapera.
For present purposes, we note the following. The parameter $E$ is
crucial in making the transition of the first order; when $E$ is non-zero
there is a second minimum at positive $\phi$ which first appears at a
temperature $T_1$, becomes stable at a temperature $T_c$ and then
becomes the true vacuum below the temperature $T_o$. Thus, we expect
that supercooling can take place as the Universe cools, the field $\phi$
getting stuck in the false vacuum state at $\phi=0$ for $T<T_c$. In a
second order transition, the parameter $E$ is zero, and there is a
smooth transition from $\phi=0$ to $\phi\neq 0$ with no supercooling.
In our context the supercooling is crucial for supplying the
non-equilibrium environment necessary for baryogenesis. The transition
from false vacuum to true vacuum proceeds by the nucleation of bubbles
of true vacuum which then expand (if large enough) to fill the universe,
baryons being produced in the vicinity of the advancing bubble walls.

In the minimal standard model, the parameter $E$, taking into account
the analysis presented above, is given by
$$E= {3\over 12\pi v_o^3} \left( 2m_W^3 + m_Z^3 \right) .\eqn\eone$$
Only bosonic fields contribute. This is understood in the present
context by the appearance of a non-analyticity in the dependence of
$I_+$ on $y^2$ coming from the minus sign preceding the exponential
in Eq.~\ipm. This non-analyticity is absent for $I_-$. It will be of
some importance to us to understand the origins of this non-analyticity.
As I will now demonstrate, it is due to infrared divergences in loop
integrals.

The cubic term in the potential may be derived in several other ways
by evaluating Feynman graphs. In order to do this, we will
find it convenient to evaluate, instead of the potential directly,
the tadpole graphs that contribute to $dV_T / d\phi$. In the minimal
standard model at one loop, we then find
$$\eqalign{{dV_T \over d\phi} &= \sum_{i,{\rm bosons}}
(h_i \phi) T \sum_n \intk {1\over\vec{k}^2 +\omega_n^2 +m_i^2(\phi)}\crr
&\quad- \sum_{i,{\rm fermions}} T \sum_n \intk
\tr \left[ h_{f,i} \; P_i (\vec{k},\omega_n,\phi) \right] \cr
}\eqn\tadI$$
where the $h_i$ are the couplings of $\phi$ to the various particles
and $P_i$ is the Euclidean fermion propagator. The trace gives a
factor of $m_i(\phi) = h_{f,i}\phi$ for the fermion contribution.
Re-writing $h_i=h_{f,i}^2$ for the fermions, we obtain
$$\eqalign{{dV_T\over d\phi} &= \sum_{i,{\rm bosons}} (h_i \phi) T
\intk {1\over \vec{k}^2 + m_i^2(\phi)} \crr
&\quad +\sum_i (h_i \phi) T \sum_{\omega_n\neq 0} \intk
 {1\over \vec{k}^2 + \omega_n^2 + m_i^2 (\phi)} .\cr}\eqn\tadII$$
I have separated off the $\omega=0$ terms as these have
different behaviour for small $\phi$. Indeed we can perform the
above integrations by dimensional regularization and the
frequency sums can be zeta-function regulated to obtain, for small
$m_i(\phi)/\pi T$, the following:
$${dV_T\over d\phi} = -{1\over 4\pi} \sum_{i,{\rm bosons}}
h_i T\phi m_i(\phi) + \sum_i c_i h_i T^2\phi + \ldots \eqn\tadIII$$
where the $c_i$ are $\phi$-independent constants. We note that the
first term, which came entirely from the zero frequency mode of the
bosons,
integrates to give a cubic term in $V_T$, whereas all of the other
modes give corrections to quadratic (and higher) terms in $\phi$, and
in particular, are {\em analytic} in $\phi^2/T^2$. Thus the cubic
term is understood as arising from the infrared region of 4-momentum
space.

Now, if we go on to powercount higher loop graphs, we find that the
infrared behaviour found at the one loop level worsens. Indeed, one finds
that there is an effective expansion parameter that is of order
$g^2 T^2 /m^2(\phi) \simeq (T/\phi)^2$, where $g$ is some generic
coupling
constant; for concreteness, we take it to be the SU(2) gauge coupling.
The appearance of this expansion parameter means that perturbation
theory breaks down at $\phi$ less than or of the order of $T$. This
is clearly a disaster for an examination of the phase transition,
where everything typically happens in this regime.\refmark\washout\
In what follows,
we will consider improvements of perturbation theory obtained by
attempting to take into account all of these infrared divergent
contributions in some consistent way. To begin, we will assume that
the coupling $g$ is sufficiently small that perturbation theory
using it as an expansion parameter makes sense. In contrast to
ordinary 4-dimensional field theory, we will see that one does
not here have the luxury of factors of $4\pi$ in the expansion
parameter. For these reasons, much of the powercounting arguments
supplied below will be merely formal arguments valid for small $g$;
an accurate numerical analysis awaits the future.\foot{In particular,
it has been claimed that (light) Higgs bosons may induce numerically
large corrections. See Refs.~\dave.}

\section{Infrared Improvements}

The fact that higher loop graphs are important near the phase
transition was first noted by Weinberg\refmark\Wein\ and was also
discussed in Refs.~{\kirli,\higher}. Indeed in
order that a theory develop a second minimum at low temperatures,
it must be that the perturbative expansion breaks down because the
second minimum is `non-perturbatively far away' from the original
vacuum state. More recently, this has been emphasized in the context
of the electroweak theory by several authors. In Ref.~\hsu, Brahm
and Hsu attempted to take into account some of the higher order graphs
but found a large {\it negative} linear term $\sim -g^3 T^3 \phi$,
leading to the conclusion that the transition is at most very weakly
first order. In a similar analysis, Shaposhnikov\refmark\shap\ found a
large {\it positive} linear term $\sim g^4 T^3 \phi$, leading him to
conclude that the transition could be very strongly first order. In
work coincident with our own, Carrington\refmark\carr\ found
corrections to the cubic term, leading her to initially conclude that
the transition is more strongly first order. All of these results have
now been understood to be incorrect, as I will explain in detail below.
In particular I will show that linear terms never arise, but that the
cubic term is modified in an important way.

\FIG\fone{Infrared divergent ring diagrams.}

Let us now look at higher loop graphs in more detail. The graphs of
Fig.~1 where the large loop is at $\omega=0$ and the smaller loops
are at $\omega\neq 0$ are the most infrared divergent graphs.
These have been referred to in the literature as `daisy' or `ring'
diagrams. If we powercount the graphs for $m(\phi )\ll \pi T$, we
find that they are proportional to
$$V_{{\rm ring},p} \sim \left( {g^2
T^2\over m^2(\phi)} \right)^p \sim \left( {T\over\phi} \right)^{2p}
\eqn\powerI$$
and thus are not formally suppressed by powers of
coupling constant. To proceed, we must resum perturbation theory in
some way. In the following sections, I will discuss in detail two
methods of understanding this resummation.

\bigskip
\noindent{\it 1.2.1~~Schwinger-Dyson Approach}
\smallskip
One notes that the diagrams in Fig.~\fone\ correspond to insertions of
one-loop propagator corrections on the large loop. This leads us to
believe that perturbation theory may be improved if we can take into
account
these loop corrections to the propagators. In an idea outlined in the
book by Kapusta,\refmark\kap\ and stressed by Carrington\refmark\carr\
in the context of the Standard Model, one extremizes the effective
action, not just with respect to the vacuum value of the Higgs field,
but also with respect to the polarization tensors,
$\langle\Pi (\omega_n,\vec{k})\rangle$. Thus we
go one step beyond the mean field approximation. To do this one writes a
shifted propagator
$$\eqalign{%
{\cal D}_{0,n}^{-1}(\vec{k}) &= \;\vec{k}^2 +\omega_n^2 + m_o^2 \crr
{\cal D}_n^{-1}(\vec{k}) &= {\cal D}_{0,n}^{-1}(\vec{k})
+ \Pi_n(\omega_n,\vec{k}) \cr}\eqn\sdprop$$
where we for simplicity give explicit formulae for a scalar boson only.
The result can be immediately generalized, and indeed we will do so
later for the Standard Model. We now add zero to the
Lagrangian in the following way:
$${\cal L} ({\cal D}_0) = {\cal L} ({\cal D}) - \shaf T\sum_n \phi_{-n}
\Pi_n \phi_n .\eqn\sdlagr$$
We are to do perturbation theory with the first term, and treat the
last term as a counterterm. We can then compute the one-loop effective
potential, which looks like\refmark\kap
$$V(\phi)=V_{tree}(\phi) - \shaf T\sum_n\intk\left[ \ln (T^2
{\cal D}_n) - \Pi_n {\cal D}_n\right] + \sum_{\ell=2}^{\infty}
V_{\ell} (\phi,{\cal D}) + {\rm subtr} .\eqn\sdpot$$
Note the important correction term to the one-loop logarithm. The
integer $\ell$ counts loops, and one should note that these higher
loop diagrams involve corrected propagators.
The self-energy is obtained self-consistently by requiring that it
extremize the potential. An arbitrary variation of this effective
potential can be written
$$\eqalign{\delta V &= \dph{V_{tree}}\delta\phi
 -{T\over 2}\sum_n\intk\left[ \left(
\Pi_n(\vec{k}) - 2\sum_{\ell=2}^\infty {\delta V_\ell\over\delta{\cal
D}_n (\vec{k})}\right) \delta{\cal D}_n (\vec{k})\right] \crr
&\quad +\left[ {T\over 2}\sum_n\intk\dph{{\cal D}^{-1}_{0,n}(\vec{k})}
{\cal D}_n (\vec{k}) + \sum_{\ell=2}^\infty\dph{V_\ell}\right]
\delta\phi , \cr}\eqn\sdvar$$
where $\phi$ and $\Pi$ have been varied independently.
Thus the requirement on the self-energy turns out to be just the
Schwinger-Dyson equation
$$\Pi_n = 2\sum_{\ell=2}^{\infty} {\delta V_\ell\over\delta {\cal
D}_n},\eqn\sdsd$$
\FIG\ftwo{Schwinger-Dyson equation.}
given diagrammatically in Fig.~\ftwo.

\FIG\fthree{
Tadpole graphs through two-loop order in Schwinger-Dyson approach.}
The variation with respect to $\phi$ gives us the equation for the
tadpole, shown in Fig.~\fthree, where the internal lines are {\it full}
propagators, and the vertices are uncorrected.
To show that the perturbative series has indeed been improved, we can
evaluate, again in just the simple scalar theory (we return to the
realistic case presently) this expression at one loop; we find
$$\dph{V} \sim \lambda\phi \intk {1\over \vec{k}^2 + m_o^2(\phi) +
\Pi_o(\vec{k})}\eqn\sdtad$$
where we have isolated the zero-frequency contribution as above.
Taking the limit $k\rightarrow 0$ within $\Pi$, we obtain
$$\dph{V} \sim -\lambda\phi T\left[ m_o^2(\phi) + \Pi_o(0)\right]^{1/2}$$
and hence a `cubic' term
$$V\sim -{T\over 12\pi}\left[ m_o^2(\phi) +
\Pi_o(0)\right]^{3/2} +\dots\; .\eqn\cubic$$
As long as $\Pi_o(0)$ is not
singular in the limit $\phi\rightarrow 0$ (it isn't), we obtain no linear
terms in $\phi$. Furthermore, if $\Pi_o(0)$ is large compared with
$m_o^2$
(true for some region of small $\phi$), the cubic term is reduced to
essentially zero. This will have important consequences later for the
electroweak potential.  In the scalar theory, Eq.~\sdsd\ gives us
at lowest order
$$\eqalign{\Pi_0(\vec{q}\rightarrow 0) &= 3\lambda T^2\left\{
\sum_n \intk {\cal D}_n(\vec{k})
\left[1 + 6\lambda\phi^2 {\cal D}_{-n} (\vec{q}-\vec{k}) \right] \right\}
\crr         &= \quar \lambda T^2 + \dots\; .\cr}\eqn\pisc$$
The terms in the ellipsis involve corrections of order $m/T$, and
include dependence on $\Pi_n$. The leading term, of order
$\lambda T^2$, is unambiguous, and we see that the would-be cubic
term is removed for small $\phi$.


We now consider the Standard Model. In this case, we run into
additional problems: in a gauge theory, the `magnetic mass' vanishes.
That is, if we compute $\Pi_o(k\rightarrow 0)$ for the transverse
gauge bosons, we find
there is no term of order $g^2 T^2$. We will see in the following
sections that this fact leads to an as yet insurmountable difficulty
with the perturbative expansion, at least in some range of $\phi$, and
has a direct analogue in finite temperature QCD.

Before discussing this further let us note the changes to the
electroweak potential coming from the resummation. The top quark
self-energy is $\Pi_{\rm top} \simeq g_s^2 T^2 /6$, but being a fermion
it has no $\omega = 0$ mode, and so does not contribute to the infrared
problem. Higher order graphs lead to numerically small corrections to
the $\phi^2$ and $\phi^4$ terms. To discuss the gauge bosons, let us
work in Coulomb gauge. In this gauge, we have propagators that mix the
transverse gauge fields with the Coulomb mode and the Goldstone mode.
At $\omega=0$, these decouple and we find that the Coulomb and
Goldstone mode self-energies have a term of order $g^2 T^2$, as does
the physical Higgs, but the two transverse modes have no such term.
This is as we discussed above. Referring to the result of Eq.~\cubic ,
we write the corrected cubic term from the $W^{\pm}$ as
$$V_{\rm cubic} = -{2\over 12\pi} T \left[ (m_W^2 + m_D^2)^{3/2} +
2(m_W^2)^{3/2} \right]  \eqn\cubicn$$
where $m_D = \sqrt{\Pi_o (k\to 0)}\sim gT$ is the Debye mass. There is
also a similar contribution from the $Z^0$. As we will see later,
this is reliable for $\phi\gsim gT$, and thus we see that the cubic
term has been reduced by a factor
of $2/3$, as the first term in the brackets of Eq.~\cubicn\ gives
(small) corrections to the $\phi^2$ and $\phi^4$ terms. This has
important consequences for baryogenesis. We note that the position of
the minimum is now given by
$$\left. {\phi\over T}\right|_{\rm min} \simeq {2\over\lambda_\epsilon}
\left( {2\over 3} E \right) (1+\epsilon)$$
where $\epsilon = (T_c^2 - T^2)/(T_c^2 - T_o^2)$ gives the
temperature of nucleation. The position of the minimum, for a given
value of $T$ has been moved in towards the origin and hence the
transition is less strongly first order (the height of the barrier is
correspondingly lower). The bound on the Higgs boson mass, found by
requiring that sphalerons not wash out\refmark\washout\ the baryon
asymmetry after the transition, is expected to be lowered by a similar
factor, down to about 40 GeV. We conclude from this analysis that the
minimal standard model is incapable of producing sufficient $n_B /s$
(regardless of the amount of {\it CP} violation) because of the LEP
bound\refmark\lep\ of $m_H \gsim 57$ GeV.

This bound is easily avoided however by extending the model. A
multi-Higgs model (two or more doublets) is particularly attractive here,
because it can easily avoid the mass bound\refmark{\bks,\tztwo} and
there is {\it CP} violation at a (nearly) sufficient level.
Anderson and Hall\refmark\andhall\
have also noted that adding singlet scalars can modify the dependence
of the Higgs mass on the self-coupling in such a way that the bound is
avoided. As well as raising the theoretical bound on the Higgs mass, an
extended model does not suffer from such a restrictive laboratory
bound as does the minimal standard model (because of, \eg , $\tan\beta$
dependence, \etc ).
\bigskip
\noindent{\it 1.2.2~~Infrared Effective Action}
\smallskip
It is also useful to understand these aspects of the electroweak
potential in an alternative way. The idea here builds on the fact that
the interesting behaviour comes from the infrared region.  Since, at
weak coupling, we
are interested in energy scales much less than $\pi T$,
one can first integrate out all modes of the various fields with
non-zero Matsubara frequency, thereby obtaining
a three-dimensional effective action for the zero-frequency bosonic
fields. As we will see below this technique automatically sums the
troublesome graphs, and automatically gives the correct combinatorics;
we are able to clearly see the absence of linear terms in the
potential.

Integrating out all of the heavy modes corresponds to evaluating all
Feynman diagrams containing loops of heavy fields, with an arbitrary
number of external light legs. Thus we obtain a three dimensional
effective action involving an infinite number of operators. In
particular, the quadratic terms pick up important contributions. In a
scalar theory, this action is given by\foot{We have used dimensional
regularization.}
$$\eqalign{\delta {\cal L}_{\rm QUAD} &= \shaf \int d^3x \phi_o^2
\left( 3\lambda T \sum_{n\neq 0} \intk {1\over\vec{k}^2
+\omega^2_n + m^2} \right) \crr
&= \shaf \int d^3x \phi_o^2 \left\{-{3\lambda T^2\over 2}
\sum_{n\neq 0} \left[ n^2 +
\left({m\over 2\pi T}\right)^2 \right]^{1/2} \right\}\crr
&= \shaf \int d^3x \phi_o^2
\left\{-3\lambda T^2 [\zeta(-1) + \ldots ] \right\}\cr}\eqn\mass$$
The quadratic part of the action thus has a term of order $\lambda
T^2$ and all corrections to this are {\em analytic} in $(m/T)^2$,
because the $n=0$ contribution is absent. The contribution of the $n=0$
mode corresponds to loop graphs in the infrared effective theory.

\FIG\ffour{One-loop diagram in infrared effective theory, showing
expansion in terms of ring diagrams.}

Having integrated out the heavy modes (at one loop level), we can now
compute the one-loop effective potential in the effective theory. This
is just the determinant of Gaussian fluctuations in the low energy
theory,
and corresponds to the bubble graph of Fig.~\ffour; if one writes this
out in terms of the heavy lines that have been integrated out (Fig.~%
\ffour ), we see that we have reproduced the sum of the dangerous
ring diagrams. One can easily check that the cubic term is correctly
reproduced, as in Eqs.~\cubic\ and \cubicn.

In the case of the Standard Model at one loop, one finds results similar
to Eq.~\mass\ for the Coulomb and scalar lines and thus one has the Debye
mass corrections of order $g^2T^2$ for these fields, whereas the
quadratic term for the transverse gauge bosons goes to zero with
$\vec k$ and $\phi$, as discussed above. These ``mass corrections'' are
analytic in $|\phi|^2$; in particular, they do {\em not} contain linear
terms in $\phi$, and thus there are no linear terms in the effective
potential, at least at this order. In principal, we should go on to
examine higher loop graphs. In the next section, we will discuss the
improvements that have been made to perturbation theory by summing the
ring diagrams. In the present context, however, we can easily understand
the absence of the linear terms at the two-loop level as well. The
dangerous graphs are those with a transverse gauge boson in one loop
and some other field with a Debye mass\foot{Recall that the loops in
the effective theory involve only $\omega=0$ modes; the loop momenta
are cut off at the Debye mass.} in the other loop. Individual graphs
are badly behaved, that is, they apparently give linear terms, but the
sum of all such two-loop graphs is not so singular. This can be
understood by thinking of the loop containing fields with a Debye mass
as a one-loop correction to the propagator of the transverse gauge boson,
\ie, by an
insertion of the polarization tensor. This has been calculated at
one-loop level and contains a momentum factor in the numerator which
softens the infrared divergence of the two-loop
graphs.\refmark\ourpapera\ Thus one sees
that there are no linear terms in the effective potential through order
$g^4\phi T^3$.

Additional higher loop graphs are potentially troublesome; these are
known in the literature by the name ``superdaisy'' graphs. In the
present context these are taken care of by further integrating out
modes. Recall that we have integrated down to a momentum scale of order
$\pi T$. In this theory, we have fields with masses of order $gT$, and
fields which are lighter. We can consider moving the cutoff lower, below
$gT$, whereupon the Coulomb and Goldstone modes are integrated out.
As long as the temperature is not too close to $T_o$, we can also
integrate out the Higgs boson as well, leaving only the
transverse modes. With
such a low cutoff, the problems arising from the superdaisy graphs are
moot. In the next section, we will study the infrared behaviour of the
full theory and the validity of perturbation theory by looking at
this low
energy theory. At temperatures in the vicinity of the phase transition,
additional divergences arise because the Higgs boson is becoming light.
In what follows we assume that the Higgs boson mass is at least of
order $gT$.

\section{How good is perturbation theory now?}

We would now like to discuss the improvement of the perturbation
expansion in
a gauge theory and discuss whether or not we can reliably predict a
first-order transition. We have noted that the expansion fails because
of the lowest-order vanishing of the magnetic mass of the transverse
gauge
bosons. We expect that the associated infrared divergences will be cut
off in some way. For the purpose of the present discussion, let us
suppose that this cutoff is provided by a magnetic mass $\mmag \sim g^2
T$. Consider then the theory where we have integrated out all modes
except for the transverse gauge bosons (\ie, at scales below
$gT$).\foot{For temperatures very close to the transition, the Higgs
will also become light, and the perturbation expansion will break down.
We suppose here that we are at sufficiently high $T$ that the Higgs
may be integrated out as well.} If we powercount vacuum graphs, we
find that they behave like
$$V_{p} \sim g^6 T^4 \left( {g^2 T \over \sqrt{m_W^2 +
\mmag^2}}\right)^p.\eqn\powerIII$$
If we neglect $\mmag$, the expansion parameter
is of order $gT/\phi$, and the expansion is valid
for $\phi\gsim gT$. This range of $\phi$ is consistent with neglecting
$m_W$ in relation to $\mmag$. We can interpret this in another way. In
the vicinity of the phase transition, $\phi/T \sim E/\lambda \sim
g^3/\lambda$, and so the expansion parameter is of order $\lambda/g^3$.
We conclude from this that large $m_H$ is a potential problem for us,
at least if we want the minimum of the potential to appear in the
perturbative regime.
\FIG\ffive{Typical electroweak potential showing hypothetical
region of validity of expansion.}

The typical shape of the potential is shown in Fig.~\ffive . The exact
placement of the limit of perturbative applicability and the position
of the minimum depend crucially on numerical factors. In any case, for
sufficiently strong first order transitions (where $\phi/T \gsim 1$) we
can reliably predict the presence of a minimum away from the origin,
and we expect a first order transition to occur. Further evidence is
provided by the $\epsilon$-expansion and the absence of an infrared
fixed point.\refmark\ginsparg

We can also worry about the presence of bad behaviour at small $\phi$,
and whether or not symmetry is truly restored at high temperatures.
Here, the behaviour of higher order graphs is similar to that of
QCD.\refmark\LGPY\
There, the potential can be written, as in Eq.~\powerIII , as
$$ V = \dots + T^4 \left\{ O(g^4) + O(g^6 \ln g^2)
+ O \left[g^6 \sum_p (g^2 T/\mmag)^p\right]
 \right\}$$
where all of the last terms are formally of the same order. The
standard wisdom is that we can compute the small $O(g^4)$ and $O(g^6\ln
g^2)$ corrections, and then hope that all the rest adds up to something
small. This is an unsolved problem. In the case of the electroweak
theory, we arrive at the same result, but we have additional
information. The cutoff at $O(g^2 T)$ implies {\it polynomial} behaviour
in $\phi$ for small $\phi$ and thus we expect, for example, the quadratic
term can be written
$$V_{\rm quad} = \phi^2
\left[-\shaf\mu^2 + 2Bv_o^2 + DT^2 + O(g^4 \ln g^2)
T^2 + \sum g^4 (g^2 T/\mmag)^p\right] .$$
Even though we cannot hope to compute past $O(g^4\ln g^2)$ here, we do
not expect any bad behaviour at small $\phi$, and thus we can claim that
symmetry is restored at high temperature, regardless of infrared
divergences, provided they are cutoff in some way. Furthermore, as
above, we conclude that the transition is first order, as long as it is
sufficiently strongly so. This includes those theories capable of
producing a baryon asymmetry.

\chapter{Bubble Wall Propagation}

A detailed study of baryogenesis at the electroweak phase transition
necessarily involves an understanding of how and when the bubbles are
formed in the first order transition, as well as how they propagate.
In this talk we will give only a cursory discussion of the
former, noting only those aspects that are of relevance to the
propagation of the bubble walls.

Studies of the nucleation event\refmark{\ourpapera,\andhall} indicate
that bubbles are formed at a temperature just below $T_c$, but
sufficiently far below this that the thin-wall approximation is not in
all respects valid. Using the improved potential evaluated at a Higgs
boson mass of 35 GeV,\foot{We choose this value here and throughout this
section because it represents roughly the largest value that would be
compatible with baryogenesis. We emphasize that the minimal standard
model is being taken here as a toy model. The strength of $CP$ violation
is of course another issue of importance for baryogenesis, but of no
importance for the propagation of the bubble wall.}
one finds
that the nucleation occurs at a temperature given by
$\epsilon\sim\quar$ where $\epsilon = (T_c^2-T^2)/(T_c^2-T_o^2)$.
Because the expansion of the Universe is so slow at this temperature,
a typical bubble grows to a macroscopic size before colliding with other
bubbles. The fractional change in the temperature of the Universe
during the period of expansion is found to be roughly ${\delta T/%
T} \sim 10^{-5}$. It is thus a good approximation to ignore the
expansion of the Universe in our calculations.

In the first scenarios proposed for the formation of the asymmetry,
baryon number was produced in the bubble wall.\refmark{\barI -
\barIV,\dine} This mechanism, at best, is not terribly efficient,
because the
baryon number violating processes turn off rapidly as the scalar
field expectation value turns on. The resulting asymmetry is
sensitive to the speed and thickness of the bubble.
The most effective scenarios for electroweak
baryogenesis have the baryons produced in front of the wall, in the
symmetric phase.\refmark{\barV,\ckn}
In this picture, scattering, for example, of top quarks
from the bubble wall leads to an asymmetry in left vs.~right-handed
top quarks in a region near the wall.  This asymmetry, resulting from
an asymmetry between reflection and transmission of different quark
helicities at the wall, biases the rate of baryon number violation
in the region in front of the wall; the resulting value of
$n_b /n_{\gamma}$ can be as large as $10^{-5}$. However, the authors
of Ref.~\ckn\ assumed that the wall was rather thin, with a
thickness of order $T^{-1}$. For thicker walls, the
asymmetry goes rapidly to zero.\foot{It may be possible to construct
extended models for which this statement is relaxed.\refmark\ann }
This can easily be understood.
In order to have an asymmetry in reflection coefficients,
the top quarks must have enough energy to pass through the wall.
For $m_t \sim 120$ GeV, this means typically the energy must
be greater than about $T/2$.  If the wall is very thick compared
to this scale, the motion of the top quarks is to a good
approximation semiclassical, and the reflection coefficient is
exponentially suppressed. The analyses of other authors also exhibit
sensitivity to the wall shape and velocity.

Clearly, then, it is important to understand how the bubble
propagates after its initial formation.  A complete
description of the wall evolution is rather complicated.  We will
see, however, that in certain limits it is not too difficult
to determine how the velocity and thickness of the wall depend
on the underlying model parameters.

We begin with a history of calculations of the bubble velocity. In
the 1970's Coleman\refmark\erice\ studied the zero temperature case
and found that the bubbles quickly accelerated to the speed of light.
At high temperatures however the Universe is filled with a plasma and
one expects that it is important to consider the effects of this plasma
on the propagation of the wall. There have been a number of studies of
the kinematics of the process.\refmark{\stein,\kaj} We will assume in
this analysis that
a steady state is reached after a sufficient time such that the wall
propagates with a constant terminal velocity. In this case we can boost
to the wall frame, in which the plasma on either side of the wall is
characterized by a mean velocity $v_{{\rm in(out)}}$ and
temperature $T_{{\rm in(out)}}$.\footnote{\star}%
{The label `in' refers to the
broken phase, and `out' to the symmetric phase.}  Applying
conservation of energy-momentum to the plasma far on either side of the
wall, one obtains two equations relating the parameters above (from
$T^{zz}$ and $T^{0z}$). The result of this analysis is that essentially
two different types of propagation are possible: {\em detonations},
characterized by $v_{\rm in} < v_{\rm out}$ and $T_{\rm in} > T_{\rm
out}$, and {\em deflagrations} characterized by
$v_{\rm in} > v_{\rm out}$ and $T_{\rm in} < T_{\rm out}$.

In order to fully determine the
velocities and temperatures, one needs additional dynamical
information. One attains this by studying the interaction of the plasma
with the wall. In the present case, the interaction is provided by the
increase in mass of the various particles as they traverse the wall. It
is clear that the most important particles here are the heaviest,
namely the top quark and the $W^\pm$ and $Z^0$. The lighter degrees of
freedom essentially pass through the wall untouched; however these
degrees of freedom constitute approximately $80\%$ of the plasma and
are important in thermalizing the distributions of heavy particles.
Because of the preponderance of light species, it is natural to suppose
that the changes in temperature
and velocity across the wall are small. This assumption allows us to
write a simple equation for the mean velocity of the wall. The
smallness of the variations in velocity and temperature will be
verified {\em a posteriori}. With these assumptions, the wall velocity
is computed by summing up all of the forces on the wall and then
solving for the velocity. In our discussion, we will assume that the
velocity is non-relativistic\foot{This assumption is not operationally
crucial, but is merely convenient.}  and expand the force in small
velocity. By looking at energy-momentum conservation in the wall frame,
we can write
$$\partial_\mu T^{\mu z}_{\phi} = \sum_i F^{(i)}_z \eqn\eom$$
where the left hand side contains the (zero temperature)
energy-momentum tensor of the scalar field, and the right hand side is
the sum of forces from the $i^{th}$ species. At zero velocity, as we
will see below, the right hand side is just the difference in pressure
of the plasmas on either side of the wall, and thus the velocity
independent part of the force is just the difference in the effective
potential on either side of the wall.

There have been a number of attempts in the past to compute the wall
velocity. In Ref.~\linde\ a simple formula for the wall velocity was
given,
based on a semiclassical picture in which a species of particle
gains a large mass $M \gg T$ as it passes through the wall.
Balancing the force on the wall due to these particles with the
pressure difference between the two phases gives a relation of the
form $$v = {\Delta p \over \Delta\rho}. \eqn\lind$$
where $p$ is the pressure and $\rho$ is the internal energy.

More recently, Turok has argued\refmark{\Turok} that this type of
analysis is incorrect.  He suggests that reflection of particles
from the wall does not slow the wall at all. The crucial step in
Turok's original analysis is the assumption that equilibrium
distributions appropriate to the value of the mass are maintained
locally. Under this assumption, the force on the wall can be written
$$\eqalign{F_z &= \int dz \intk n(\vec{k},\vec{x})\; v_z {\Delta p_z
\over \Delta z}\crr
&= \int dz {\partial m^2\over \partial z} \intk {1\over 2 k_o}
n(\vec{k})\crr
&= \;{\rm velocity-independent}.\cr} \eqn\turanal$$
The distribution $n$ is taken to be the boosted Bose-Einstein (or
Fermi-Dirac) distribution. The second line follows by writing the
change in momentum in terms of the change in mass as the particle
traverses the wall (using energy conservation). The last line follows
since the integral over momentum involves Lorentz-invariant quantities
only, and thus the velocity dependence may be removed by a change of
variables. With this analysis one concludes that the wall accelerates
relentlessly, at least until some other effect takes over.
This analysis is indeed correct; it is the assumption of equilibrium
that is physically unmotivated. We will see shortly that the most
important effects arise because of small, velocity-dependent
departures from equilibrium.

Before going on to the more realistic case hinted at in the last
paragraph, we will improve on the analysis of Ref.~\linde ; in the
course of doing this we will identify some of the shortcomings of this
analysis.

\section{Thin Wall Analysis}

We will now make the following assumptions: we suppose that the
individual particles of the plasma, while interacting with the wall,
do not interact among themselves. We thus may follow individual
particles as they traverse the wall and compute the force accordingly.
This assumption obviously corresponds to assuming that any relevant
mean free path is long compared with a scale set by the size of the
wall. This, it will turn out, is an implicit assumption in the estimate
of Ref.~\linde . However, we can improve on that analysis by noting
that in reality there are no particles of relevance with mass much
larger than the temperature. Depending on the momentum, a certain
fraction will be reflected from the wall, but a significant fraction is
also able to pass through the wall in either direction. If we then make
the simple assumption that the incoming particles have a roughly
thermal distribution, we can compute the total force on the wall by
carefully adding up the contributions from each type of event noted
above. The details of this calculation are given in Ref.~\ourpapera,
and we will simply quote the result here. If we write the force as
$$0 = \Delta V_T + v\;\sum_i {\cal E}_i + O(v^2)$$
we obtain, in the thin wall case,
$${\cal E}_i = \rho(o,T) - \rho(m_i,T) - {m_i^2\over4\pi^2}
\int_{m_i}^\infty E\; n_o (E) dE .\eqn\thin$$
We note that in the limit $m\rightarrow\infty$, the result of Ref.~%
\linde\ is recovered. If we expand this expression in powers of $m/T$,
we find the following result. For top quarks, we have\foot{For the sake
of this discussion, we are assuming that the expansion is valid for top
quarks.}
$${\cal E}^{thin}_{f}\simeq -{3 \phi^4\over 16\pi^2 v_o^4} m_t^4
\left[ \ln {m_t^2 \phi^2\over a_F v_o^2 T^2} -{7\over 2} \right]
+ O\left[\left({m\over T}\right)^5\right]\eqn\topthin$$
and for $W$ and $Z$ bosons, we have
$${\cal E}^{thin}_{b}\simeq {3\over\pi} ET\phi^3
+ {3\phi^4\over 64\pi^2 v_o^4} \left[ 2m_W^4 \left( \ln
{m_W^2\phi^2\over a_B v_o^2 T^2} -{7\over 2}\right) + m_Z^4 \left(
\ln{m_Z^2 \phi^2\over a_B v_o^2 T^2} -{7\over 2}\right) \right] \ .
\eqn\bosthin$$
If we ignore the top quark, the velocity turns out to be a pure number
times a simple function of $\epsilon$,
$$v\simeq {\pi\over 6}{\epsilon\over 1+\epsilon}\;\;(\sim 0.1\ {\rm for}\
m_H =35\ {\rm GeV})$$
valid for small $\epsilon$. Putting back in the top quark, we find that
the velocity, in this approximation, is diminished. For $m_H= 35$ GeV
and $m_t=120$ GeV, one finds $v\sim 0.07$.

We may now go back and check the validity of our assumptions. First,
the working assumption of non-relativistic velocities clearly holds.
Secondly, if we go back to the kinematics and compute the variations in
velocity and temperature consistent with energy-momentum conservation
we find
$$\eqalign{{\delta v\over v} \sim & + 10^{-2} \crr
{\delta T\over T} \sim & - 10^{-4} .\cr}$$
This result is clearly consistent with our assumption of slowly varying
temperature and velocity. Furthermore, the signs of these quantities
are such that we identify the process as a (weak) deflagration.

\section{Parameters of Interest}

In order to proceed, we now investigate various distance scales
relevant at this phase transition. To get an idea of the thickness of
the wall, we note that at $T_c$ there is an exact kink solution known
of the form $\phi = \phi_c/2 [1+ \tanh (2z/\delta)]$ and
$\delta$ characterizes the wall thickness. In our case, we find
$$\delta \simeq {2\sqrt{2\lambda}\over E} T^{-1} \sim 40\; T^{-1}.$$
The value quoted is for $m_H = 35$ GeV. From numerical calculations,
this width is accurate at the actual nucleation temperature as well,
and thus we see that the wall is rather thick.\foot{We make the
assumption that the thickness of the wall remains constant throughout
the expansion.}

It is useful to compare this number with the mean free
paths for various processes.  In considering the properties
of the bubble wall, the relevant mean free paths are those
for particles which interact with the wall, \ie, principally
top quarks, $W$'s and $Z$'s. The processes with the shortest
mean free paths are elastic scatterings.  These exhibit the
characteristic singularities of Coulomb scattering at small
angles.  What actually interests us, however, is the momentum
and energy transfer in these collisions.  This is a problem which
has been extensively studied, and we can borrow the relevant
results.  One finds that the momentum loss per unit length due to
scattering is in all cases much larger than that due to the wall.
This result may be understood in an alternative way.  The elastic
scattering cross section diverges at small angles in empty space.
In the plasma, we expect that this divergence is cut off, essentially
by a temperature factor. Examining the expression for the
elastic scattering cross section, one obtains an estimate for the
mean free paths of order $\ell \sim 4 \, T^{-1}$ for
quarks, and $\ell \sim 12 \, T^{-1}$ for $W$'s and $Z$'s. We will use
these estimates in what follows but it should be kept in mind that a
more complete analysis of the mean free paths is desirable.

{}From the above discussion we see that the mean free paths for
thermalization processes are typically a significant fraction of the
wall thickness.\foot{Also of interest are mean free paths for processes
that change some approximately conserved quantum number. It has been
speculated\refmark\ourpapera\ that a ``snowplow'' effect may occur in
front of the wall which may further act to decrease the wall velocity.
Estimates given elsewhere indicate that this effect is at most of
approximately the same strength as the effects discussed herein.}
We thus expect that a more realistic analysis of the
bubble wall velocity lies somewhere in between the thin-wall analysis
given in the previous section and the equilibrium analysis of Ref.~
\Turok . There have been two attempts at this in the literature. The
first\refmark\ourpapera\ builds on the thin-wall analysis, whereas the
second\refmark\lmt\ perturbs around the equilibrium picture. Both
approaches show qualitative agreement that the wall velocity is
non-relativistic. We will briefly discuss these two approaches in the
following two sections.

\section{DLHLL Thick Wall Analysis}

To get an idea of how finite elastic scattering lengths affect the
velocity of the bubble wall, we assume that particles propagate
freely over distances of order a mean free path, $\ell$. As shown in
the previous section this is typically shorter than the thickness of
the wall and so we view the bubble wall as a succession of slices with
thickness of order $\ell$, and for each of these we repeat
the thin wall analysis. Thus we make the simple assumption that the
distributions are given roughly by equilibrium distributions
appropriate to a given point within the wall, and then follow that
distribution over a length of order $\ell$. On general grounds, one
expects that the result may be written in the following form:
$${\cal E} = {\cal S}_b {\cal E}_b^{thin} + {\cal S}_f {\cal
E}_f^{thin} , $$
where ${\cal S}$ are suppression factors dependent on $\ell$. One
expects them to have several limiting behaviours. When
$\ell/\delta\rightarrow 0$ they should approach zero, reproducing the
equilibrium analysis in which the velocity dependence goes away. When
$\delta<\ell$, the thin wall analysis is recovered. Numerically, one
finds that the suppression factors are not very sensitive to $m_t$
and $m_H$, and are well fit by
$${\cal S}\simeq {\sqrt{2} (\ell/\delta)\over
\sqrt{1+ (\ell/\delta)^2}}.$$
Using the values quoted above for the mean
free paths, the equations for ${\cal E}$ from
the previous section and assuming $m_H\sim 35$ GeV ($\delta\sim
40\, T^{-1}$) and $m_t\sim 120$ GeV, we find a velocity of about
$v \sim 0.2 \ .$
\FIG\fsix{Plot of velocity versus $m_t$ for several values of $m_H$
for thick and thin wall approximations.}
Fig.~\fsix\ illustrates the velocity for a range of top masses,
for two values of the Higgs boson mass. Also included on this plot for
comparison is the thin wall result for these Higgs boson masses.

\section{LMT Thick Wall Analysis}

An alternative analysis of the thick wall velocity has recently been
performed in Ref.~\lmt . These authors study linear perturbations of
the distribution function $n$, determined by the relativistic Boltzmann
equation
$$p_z \partial_z n + m\gamma F_z {\partial\over\partial_{k_z}} n =
C(n) .\eqn\Boltz$$
Here the force term is given by the change in mass of the particles in
the wall $F_z =-{1/2k_o} \partial_z m^2$.

Let us now make the relaxation time approximation, whereby the
collision term is replaced by $C=\delta n / \ell$ where $\ell$ is again
the relevant mean free path. One can then solve for $\delta n = n-n_o$
in terms
of the equilibrium distribution function and the velocity of the wall.
Balancing the forces on the wall one obtains the result of Turok, with
an additional velocity dependent term
$$\Delta V_T = \gamma v {\ell\over\delta}{E\over\pi}\;T\phi^3 .$$
Solving for the velocity, one obtains
$$\gamma v \sim c\; {\pi\over 6}\;{\delta\over\ell}\;
{\epsilon\over 1+\epsilon} ,$$
where $c$ is of order unity, a result which obviously agrees very
well with our result for very small $\ell/\delta$.

The authors of Ref.~\lmt\ however perform a more detailed analysis,
whereby they solve Eq.~\Boltz\ directly without making the relaxation
time approximation. The result of this analysis differs from the above
in some respects, but within the range of parameters of interest agrees
qualitatively with the above result. One must conclude from these
analyses that the simple analysis of Ref.~\ourpapera\ is equivalent to
the relaxation time approximation, but extends the analysis beyond very
small $\ell/\delta$. The conclusion in all cases is that the velocity
of the bubble walls is non-relativistic.

\chapter{Conclusions and Acknowledgments}

In this talk, we have discussed the recently improved understanding of
the effective potential describing the electroweak phase transition of
the minimal standard model, and outlined several ways of understanding
the resummation necessary to improve perturbation theory. We have also
discussed recent calculations of bubble wall properties, understood how
they relate to one another, and noted that
all calculations are in qualitative agreement that the wall velocity
is non-relativistic.

I would like to thank Patrick Huet, Michael Dine and Andrei Linde for
their collaboration on these projects and for many fruitful and
interesting discussions. In addition, Andrei Linde was helpful in
clarifying the historical record.
I also wish to thank L. McLerran for discussion of his work.

\endpage
\refout
\figout
\end